\documentclass[letterpaper]{nature}

\usepackage{color}
\usepackage{url}
\usepackage{verbatim}
\usepackage{multirow}
\usepackage{xspace}
\usepackage{graphicx}
\usepackage{amsmath} 
\usepackage{bbm}
\usepackage{amssymb}
\usepackage{booktabs} 
\usepackage{times} 
\usepackage{lscape}
\usepackage[left=1in,right=1in,bottom=1.1in,top=1in]{geometry}
\usepackage{enumitem}
\usepackage{lscape}
\usepackage[table,xcdraw]{xcolor}
\usepackage[normalem]{ulem}
\usepackage{xcolor}
\usepackage[innercaption]{sidecap}
\usepackage{lineno}
\usepackage[colorlinks,citecolor=blue,urlcolor=magenta]{hyperref}
\usepackage[font=small,labelfont=bf]{caption}
\usepackage[CaptionAfterwards]{fltpage}

\newcommand{\hide}[1]{}

\let\oldnl\nl
\newcommand{\nonl}{\renewcommand{\nl}{\let\nl\oldnl}} 
\usepackage[normalem]{ulem}

\graphicspath{{./FIG/}}

\title{\center\Large
\begin{center}
OncoNet: Weakly Supervised Siamese Network to automate cancer treatment response assessment between longitudinal FDG PET/CT examinations
\end{center}}

\author
{\begin{center}
Anirudh Joshi,$^{1,4, \ast}$ Sabri Eyuboglu$^{1,4 \ast}$, Shih-Cheng Huang$^{3,4 \ast}$, Jared Dunnmon $^{1}$, Arjun Soin$^{1,4}$, Guido Davidzon $^{2}$, Akshay Chaudhari $^{2,3 \ast}$, Matthew P Lungren $^{2,3,4 \ast}$\\[1mm]
\normalsize{$^{1}\;$Department of Computer Science, Stanford University, Stanford, CA, USA} \\[1mm]
\normalsize{$^{2}\;$Department of Radiology, Stanford University, Stanford, CA, USA} \\[1mm]
\normalsize{$^{3}\;$Department of Biomedical Data Science, Stanford University, Stanford, CA, USA} \\[1mm]

\normalsize{$^{4}\;$Center for Artificial Intelligence in Medicine and Imaging, Stanford University, Stanford, CA, USA} \\[1mm]

\normalsize{$^\ast$Equal Contribution. $\ddag$Corresponding author. Email: anirudhjoshi@cs.stanford.edu}
\end{center}
}

\begin{document}

\maketitle


\begin{abstract}
{\spacing{1.4}
FDG PET/CT imaging is a resource intensive examination critical for managing malignant disease and is particularly important for longitudinal assessment during therapy.  Approaches to automate longtudinal analysis present many challenges including lack of available longitudinal datasets, managing complex large multimodal imaging examinations, and need for detailed annotations for traditional supervised machine learning. In this work we develop OncoNet, novel machine learning algorithm that assesses treatment response from a 1,954 pairs of sequential FDG PET/CT exams through weak supervision using the standard uptake values (SUVmax) in associated radiology reports.  OncoNet demonstrates an AUROC of 0.86 and 0.84 on internal and external institution test sets respectively for determination of change between scans while also showing strong agreement to clinical scoring systems with a kappa score of 0.8. We also curated a dataset of 1,954 paired FDG PET/CT exams designed for response assessment for the broader machine learning in healthcare research community. Automated assessment of radiographic response from FDG PET/CT with OncoNet could provide clinicians with a valuable tool to rapidly and consistently interpret change over time in longitudinal multi-modal imaging exams.
}
\end{abstract}

{\spacing{1.4}
\section{Introduction}
Cancer is one of the leading causes of death worldwide and accurate diagnosis, staging and restaging are essential to  optimize therapeutic management. Advanced medical imaging techniques such as  positron emission tomography (PET) coupled  with computed tomography (CT) are integral for clinical assessment of cancer diagnosis, and assessment of treatment response. PET is the most sensitive non-invasive imaging modality capable of detecting picomolar amounts of radiolabeled sugar molecules trapped in cancer cells while CT provides high tissue resolution for precise localization.   The clinical interpretation of PET/CT scans involves synthesizing multiple data sources: clinical information, the metabolic findings from PET  and the anatomic information from  CT. 

In clinical practice, radiologists and nuclear medicine physicians must interpret consecutive PET/CT examinations to determine whether a cancer patient receiving treatment is appropriately responding to therapy. They do this by measuring whether the amount of malignant tissue is decreasing, unchanged or increasing across exams. This process is chiefly qualitative and sometimes subjective, but oncology treatment planning increasingly demands standardized, quantitative data \cite{ding2014pet}. Further, the process is extremely labor intensive and time-consuming and a stark rise in utilization of FDG-PET/CT imaging suggest that automation technologies would be of high impact in clinical oncologic imaging workflows.  

Deep learning approaches have produced state of the art results for automated interpretation of various medical imaging modalities \cite{irvin2019chexpert, huang2020penet, park2019deep}. Prior work in deep learning for PET/CT has demonstrated the ability of automated systems to detect and estimate locations of abnormalities in individual PET/CT imaging exams \cite{eyuboglu2021multi}. However, as discussed above, there is a pressing need for automated methods capable of comparing across consecutive oncologic imaging studies and estimating changes in disease burden.  If successful, automation of this clinically important task could improve routine clinical oncologic imaging workflows, enhance standardized quantification of imaging biomarkers for oncologic therapy trials, and contribute to operationalization of   communications regarding  response to therapy to referring clinicians and patients.

Existing approaches that compare consecutive medical images over time have been validated on 2D imaging modalities such as radiography and retinal fundus imaging \cite{arcadu2019deep, li2020siamese}. However, applying these techniques to PET/CT presents significant new challenges such as: (1) a single PET/CT exam is composed of hundreds of PET and CT image slices, so training a model to compare multiple complex PET/CT examinations is methodologically  challenging and computationally expensive, (2) established scoring systems for measuring longitudinal changes are subjective, so human readers often produce inconsistent scores, (3) clinical information depicting longitudinal changes in consecutive PET/CT imaging studies are reported in narrative text reports that make it challenging to extract meaningful labels for deep learning training in large datasets, and (4) existing work in PET/CT are limited by small reported experimental datasets or reliance on phantoms which limits scientific advancement. 

The purpose of this study is to model the task of longitudinal treatment response assesment on volumetric multi-modality oncologic PET/CT imaging examinations for automatically determining disease progression in pairs of FDG-PET/CT studies obtained before and after treatment. 

Our contributions include 1) a dataset of 1,954 labeled PET/CT imaging examinations using detailed clinical reports for multi-modal model development 2) OncoNet, a 3D deep learning model that performs treatment response assessment at AUROC 0.85 [0.7-0.95] on the internal test set 3) external validation of OncoNet on multi institutional data 4) comparison of OncoNet predictions to clinical scoring produced by a board certified radiologist. We hope our annotated dataset and methodological contributions serve as a stepping stone towards achieving automated, quantitative oncologic imaging evaluation over time with broad implications for cancer care.

}

{\spacing{1.4}
\section{Methods}

\subsection{Data}

\textbf{Training Data: } This retrospective study was approved by our institutional review board with waived patient consent. Our training dataset consists of  2572 deidentified FDG PET/CT scans from 656 patients leading to 1954 paired longitudinal scans. The exams were administered at Stanford hospital between 2003 and 2010. Each FDG-PET/CT exam also included a free-text unstructured report compiled by board-certified  radiologist at the time of the examination. The image resolution of the FDG-PET scan was 128 × 128 pixel, while that of the CT scan, was 512 × 512 pixels. 

The exams were split into training (1888), validation (33), and test sets (33). The validation and test sets were sampled at random with uniform probability to capture the class distribution expected in a clinical setting. The exams were split by patient, ensuring that there was no patient overlap between the training, validation, and test sets. The validation set was used to evaluate prototypes during model development, to tune hyperparameters using average AUROC, and for early stopping during model training. All reported metrics were computed on the test set, unless otherwise specified. 

The training set annotations were extracted using a rules based heuristic on the radiology reports.  Radiology reports contain Findings sections which detail sections of the scan (Head and Neck, Thorax, Abdomen) along with information on lesions identified, measurements and SUV values. We propose to use the maximum SUVs (SUVmax) recorded for lesions in the thorax before and after treatment as weak supervision for OncoNet. If the difference in SUVmax values between the scans is greater than 25\% or less than -25\%, the label is considered tumor progression and resolution  respectively and is stable if between -25\% and 25\%, in accordance to the Lugano 2014 criteria for tumor evaluation, assessment, and response prediction using 18F-FDG PET/CT \cite{van2017lugano}.
The validation and test set SUV annotations are determined by a board certified radiologist reviewing the radiology reports. For each exam, the radiologist assigned a single SUV score that corresponded to the most metabolically active lesion reported. The categorical label was determined as above using the Lugano 2014 criteria. 

\textbf{Internal Test Set: }
Our thorax test set consisted of 46 scans from 13 patients, leading to 33 paired longitudinal scans. The exams were sampled randomly from the original dataset and contain 11 pairs of each of the three classes.

\textbf{External Test Set: } For our external validation we used a public dataset (ACRIN 6668) from The Cancer Imaging Archive which contains studies from 242 patients. The study was conducted as a multicenter trial with the goal of determining whether PET SUV uptake in non small cell lung carcinoma (NSCLC) was a useful predictor of long-term clinical outcome (survival) after definitive chemoradiotherapy. Using the metadata where SUVmax was recorded for the thorax led to a subset of 60 patients. From this subset we filtered out scans where the PET and CT did not take place at the same date or had different number of slices. For the subset of the dataset that we selected, 18 patients saw significant reductions in SUVmax of the hottest lesion categorizing them all as tumor resolution, 1 patient was characterized as tumor progression and 1 patient was characterized as stable. All of the exams demonstrated an improvement based on a 25\% reduction in SUVmax values post treatment.

For the longitudinal pertubation experiment, the baseline and follow up scans are reversed and provided as input to the model with the flipped label. 

\subsection{Network Architecture}

OncoNet consists of three main components; Encoder, Decoder, Classifier Head. 
The Encoder is formed using the Inflated Inception V1 3D convolutional neural network (I3D) pretrained on Kinetics using optical flow \cite{carreira2017quo}. The final classification layer was removed making the output encoding a 3 dimensional encoding of the input scan. The encoding shape is $7 * 7 * \frac{l}{6}$  where $l$ is the number of slices in the original exam. The Decoder consists of a soft attention mechanism and a linear classification layer. The soft attention is a dot product between each voxel $e_{i,j,k}$ in the encoded representation and a learned weight matrix $w$. A softmax is applied on the scores computed by the dot product and a linear combination of the voxels is computed. The intuition behind the soft attention is to place higher weight on certain voxels of the exam in a data-driven manner while determining change in tumor burden. This encoder-decoder structure has shown effectiveness in prior work on PET/CT abnormality detection \cite{eyuboglu2021multi}.

The encoder and decoder weights are shared during each forward pass for each exam in the pair of exams. The output representations from the decoder are used to compute a difference representation which is a flattened tensor of dimension (hidden size,) which is then passed into a classifier head to determine response to treatment. The classifier head is formed of two linear layers with a ReLU activation function in between. 

\begin{figure}
\includegraphics[width=\textwidth]{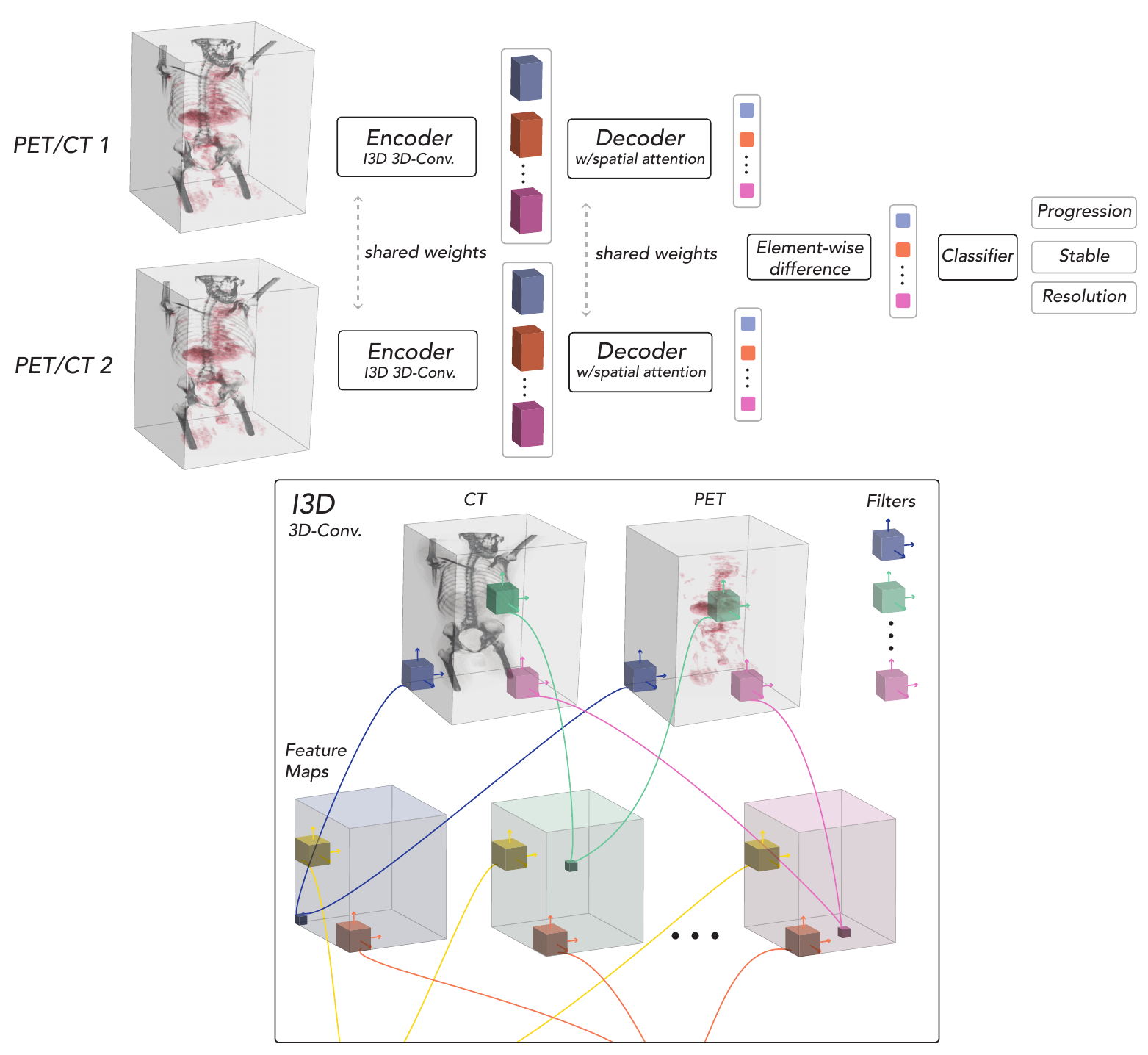}
\caption{Network architecture for OncoNet. Each PET/CT exam in a pair is run through the encoder-decoder forward pass with weights shared between the two exams in the pair. The decoder representations are diffed and the resulting representation is passed through a classifier head to determine Progression, Resolution and Stable}
\label{fig:arch}
\end{figure}

\subsection{Training Procedure}

The model is trained with a batch size of 2 for a maximum of 30 epochs using early stopping based on validation loss. The learning rate used is 0.0001 with an Adam optimizer and a step decay of the learning rate every 10 epochs by a factor of 0.1. Dropout is used as regularization during training in the classifier head. Cross entropy loss is used for supervision and the models are trained using Pytorch 1.4 on 2 NVIDIA TITAN V GPUs.

Data augmentation is also used during training to improve generalization and prevent overfitting. 3 types of data augmentation are used. We randomly crop each slice of the exam to 200 x 200 and upscale to 224 x 224, downsampling the 512 x 512 CT image and upsampling the 128 x 128 PET.  We also introduce rotations in all slices of the CT and randomly crop at 200 x 200 and rescale to 224 x 224.

\subsection{Model Evaluation}

The test set SUV annotations are determined by a board certified radiologist reviewing the radiology reports. For each exam, the radiologist assigned a single SUV score that corresponded to the most metabolically active lesion present in the scan. The categorical labels for tumor response were determined by the percentage difference in SUVmax scores for the paired scans pre and post treatment. If the percentage difference was greater than 25\%, the response was determined to be “progression”. If the percentage was less than -25\%, the response was determined to be “resolution” and values in between were determined to be “stable”.

Each model was evaluated using AUROC on the predictions compared to the test set labels AUCs were computed for each class and each region individually. F1 scores were also computed. 95\% confidence intervals were computed using 1000 bootstrap replicates for average AUROC across classes.  The models were also evaluated on an external public dataset from additional institutions in a similar manner. The SUVmax values for the external dataset were derived from the dataset metadata directly and not through radiology reports.

We visualize the model outputs using gradient based Guided Backpropagation saliency maps which compute the gradients of loss with respect to the original pixels in each 2D slice of the exam. These gradients are plotted to demonstrate the pixels that are most sensitive to the model prediction.

\subsection{Deauville Score Agreement}

A board certified radiologist reviewed each of the exams in the internal Thorax test set and assigned scores based on the Deauville criteria. The Deauville criteria is an internationally accepted scoring system that evaulates FDG avidity of a tumor mass as seen on FDG-PET. The criteria specifies a 5 point scale:
\begin{itemize}

\item Score 1: No uptake above the background
\item Score 2: Uptake $\leq$ mediastinum
\item Score 3: Uptake $>$ mediastinum but $\leq$ liver
\item Score 4: Uptake moderately increased compared to the liver at any site
\item Score 5: Uptake markedly increased compared to the liver at any site
\end{itemize}

To measure agreement we evaluate based on two stratification strategies. 1) We compare OncoNet's agreement with determination of worse/not worse by mapping the Deauville scores 1,2,3,4 to OncoNet's resolution and stable classes and the score of 5 to OncoNet's progression class. 2) We compare OncoNet's agreement with resolution and progression by mapping the resolution class to Deauville score 1,2,3 and the progression class to Deauville score 5. 
}

{\spacing{1.4}
\section{Results}
\subsection{Weakly-supervised siamese-style training enables automated treatment response predictions in the thorax}

Annotating every lesion in thousands of pairs of CT scans to compare tumor progression is a highly costly and time consuming endeavour for healthcare systems. To allow any healthcare system to train OncoNet, we leverage information recorded in readily-available radiology reports as weak supervision for treatment response. Standard Uptake Values (SUVmax) of metabolically active tumors are extracted from reports through rules based heuristics as described in methods and the differences in the SUVmax values between the scans are categorized as progression, resolution and stable. 

OncoNet uses an encoder-decoder architecture previously validated on the task of anatomically-resolved PET/CT abnormality detection \cite{eyuboglu2021multi}.   Like a siamese neural network\cite{koch2015siamese}, OncoNet computes decoder representations from two forward passes (one for each PET/CT scan in the pair). Finally, it computes the elementwise difference between the two representations and feeds it into a final classification layer. As an ablation, we study an approach that computes a difference between the PET/CT imaging exams prior to passing into the encoder-decoder network and runs a single pass through the network. This single pass approach has been studied in prior literature in automated disease progression prediction \cite{arzhaeva2010automated}.

We find that OncoNet scores 0.84 [0.95 - 0.75] AUROC on the task of longitudinal change prediction on thorax and outperforms the single pass approach by 0.14 AUROC (p $<$ 0.01). From Fig \ref{fig:perf} we see that OncoNet scores 0.93 [0.84-1.0] AUROC on Progression, 0.81 [0.56-0.98] on Resolution and 0.78 [0.59-0.95] on Stable. 95th percentile confidence intervals are computed using 1000 bootstrap replicates.

In Figure \ref{fig:tsaliency} we produce guided backpropagation saliency maps over each slice of the PET/CT to understand which voxels OncoNet's classification is most sensitive to. The top row shows an instance of diseases progression where saliency is concentrated on the new tumor in the lower left. The middle row shows a stable tumor nodule. There is no spike in saliency on the stable tumor suggesting that the model is not focused on solely abnormality detection but change in abnormality. Finally, in the bottom row we see that when there is a reduction in the disease from the previous scan, the saliency identifies the region where tumor was resolved. 

\begin{figure}
\includegraphics[width=\textwidth]{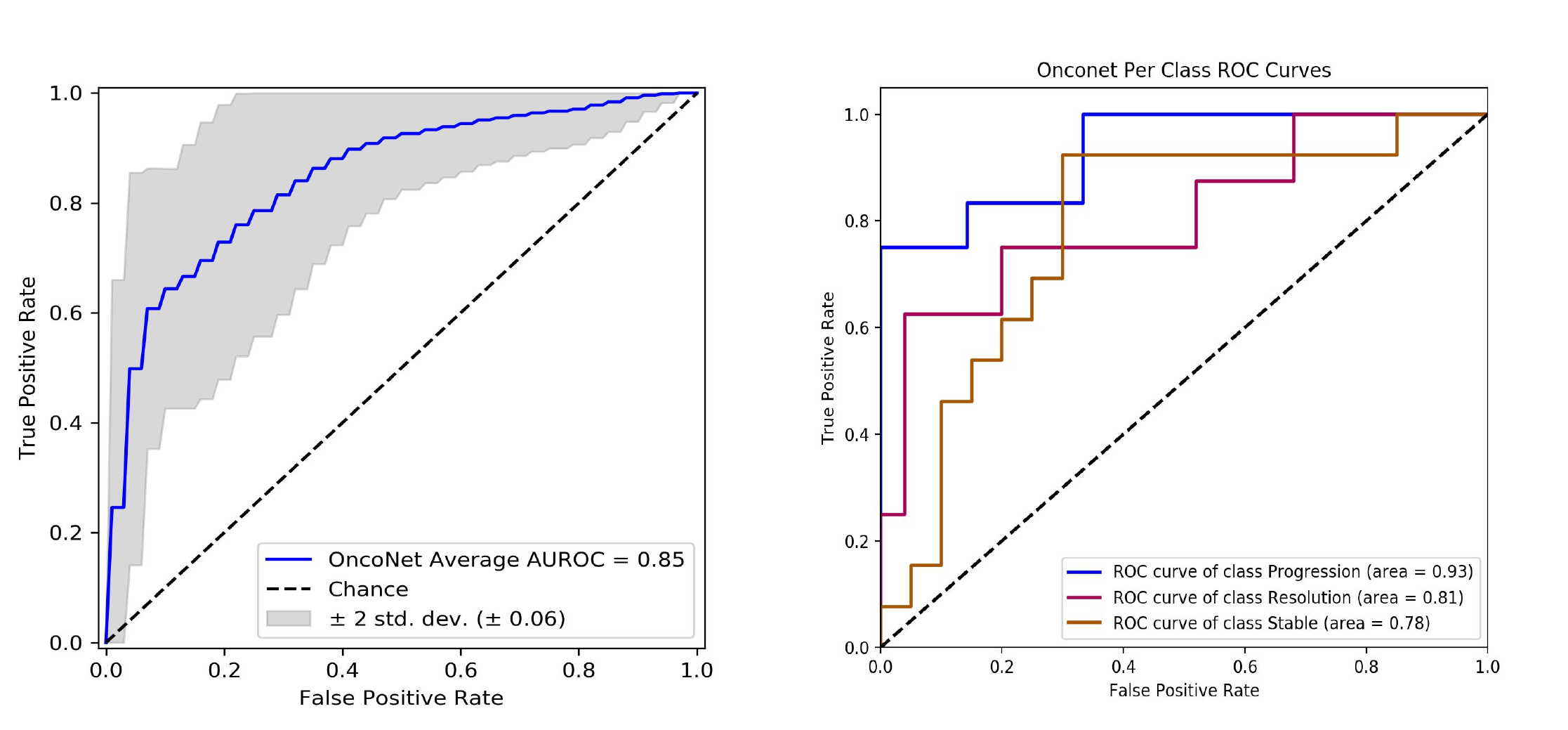}
\caption{ROC curves for OncoNet test performance. (Left) Microaveraged ROC curves over the three classes with the grey bounds indicating 2 standard deviations computed by bootstrap sampling on 1000 samples. (Right) The individual class ROC curves showing that Progression (AUROC: 0.93), Resolution (AUROC:0.81) and Stable (AUROC: 0.78) }
\label{fig:perf}
\end{figure}

\begin{center}
\begin{table}
\begin{tabular}{ | c | c|c |c| c| } 

\hline
Model & AUROC & F1 & Precision & Recall \\ 
\hline
OncoNet & 0.85 [0.7-0.95] & 0.70 [0.54-0.85] & 0.73
[0.56-0.88] & 0.70
[0.55-0.85]\\ 
\hline
Single Pass & 0.67
[0.53-0.82] & 0.54 [0.36-0.72] & 0.57
[0.37-0.76]
 & 0.55
[0.40-0.72]
 \\ 
\hline

\end{tabular}
\caption{Class averaged performance metrics along with 95\% confidence intervals computed by bootstrap sampling over 1000 samples. OncoNet significantly outperforms the single pass ablation (p$<$0.01)}
\label{table:perf}
\end{table}
\end{center}

\begin{figure}
\includegraphics[width=\textwidth]{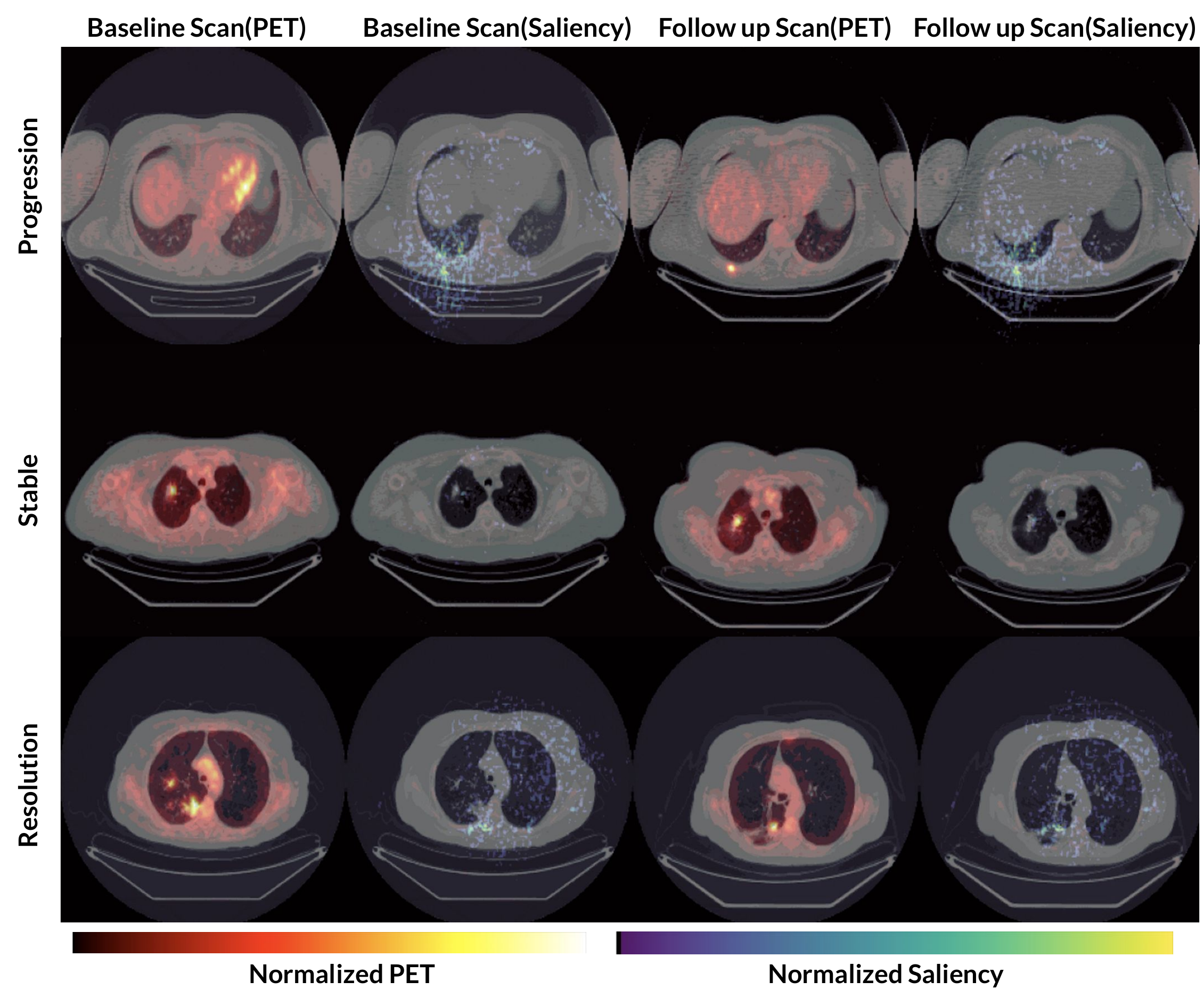}
\caption{Examples of Saliency Maps produced by OncoNet on true positives from the test set. The first and third column visualize the PET overlayed on the CT for the baseline and follow up scans. The second and the fourth column visualize the saliencies overlayed on the CT for the baseline and follow up. The top row shows the emergence of a new tumor (progression) in the bottom left part of the scan and the saliency focuses on the region of the change. In the middle row, the same tumor is seen across both scans indicating stable disease. The saliency is faint given that OncoNet focuses on change in disease and not the presence/absence of disease. In the bottom row the tumor present in the baseline regresses and the saliency focuses on the region where the tumor reduced.}
\label{fig:tsaliency}
\end{figure}

\subsection{Onconet maintains performance on data from external institutions}
We evaluated OncoNet on a public external dataset from The Cancer Imaging Archive contributed by the ACRIN Cooperative Group \cite{machtay2013prediction}. Details on inclusion criteria and the dataset are included in methods.

OncoNet maintains internal test set performance by scoring an average AUROC of 0.84 on the external test set. 18 out of 20 exams present in the external test set are tumor resolution, so we propose a method to evaluate generalization of deep learning models for longitudinal analysis by flipping the ordering of scans in the pair. We flip the progression and resolution pairs of exams resulting in a dataset of 18 progression, 1 resolution and 1 stable. We find that OncoNet scores an AUROC of 0.86 on the flipped external test set. 
We extend this evaluation to our internal test set by flipping the progression and resolution exams. We find that OncoNet scores average AUROC of 0.81 on the perturbed internal test set.

The single pass ablation performs worse on the external set compared to internal by 0.11 AUROC, however maintains performance on the flipped perturbation for both internal and external test sets.See Table 2 for complete metrics.

\begin{center}
\begin{table}
\begin{tabular}{ | c | c| c|c |c |c| c| } 

\hline
Model & Data Set & Flipped & AUROC & F1 & Precision & Recall \\ 
\hline
OncoNet & Internal & No & 0.85  & 0.70 & 0.73
& 0.70
\\ 
\hline
OncoNet & External & No & 0.84  & 0.81 & 0.91
& 0.75
\\ 
\hline
OncoNet & External & Yes & 0.86  & 0.92 & 0.70
& 0.78
\\ 
\hline
OncoNet & Internal & Yes & 0.81  & 0.61 & 0.63
& 0.64
\\ 
\hline
Single Pass & Internal & No & 0.67
 & 0.54 & 0.57
 & 0.55
 \\ 
\hline
Single Pass & External & No & 0.56
 & 0.59 & 0.89
 & 0.45
 \\ 
\hline
Single Pass & External & Yes & 0.86
 & 0.80 & 0.86
 & 0.75
 \\ 
\hline
Single Pass & Internal & Yes  & 0.67
 & 0.54 & 0.59
 & 0.54
 \\ 
\hline

\end{tabular}
\caption{Class averaged performance metrics to compare performance across internal and external test sets along with the flipped perturbation}
\end{table}
\end{center}

\subsection{OncoNet predictions correlate with the Deauville clinical scoring system}

A board certified radiologist compared each paired FDG-PET/CT scan in the test set and scored the scans based on the Deauville five point scale. The scoring system is routinely used in clinical practice to quantify treatment response in FDG PET/CT. 

A score of 1,2,3 indicates that there was response to treatment with a granular breakdown provided in the methods. A score of 4 could indicate a partial response where the metabolic activity has either remained constant or reduced but not below the level in the liver. A score of 5 indicates a new tumor or increased metabolic activity compared to the previous scan.

We study whether OncoNet’s predictions agree with the radiologist based on two levels of stratification. (1) In clinical practice stratifying a patient in categories 1,2,3,4 vs 5 is important for the radiologist to determine whether patients are either getting worse or not. We evaluate agreement with 5 by using the label corresponding to disease progression. Labels corresponding to resolved or stable would agree with categories 1,2,3,4. 
(2) Since there isn’t a direct one to one mapping between the “stable” model class and the Deauville scoring system we also evaluate another stratification to study
whether using the outputs of “resolution” and “progression” correlate to categories 1,2,3 and category 5 respectively. We selected all the exams where the model predicts “resolved” and “progressed” and of those selected the exams that received scores of 1,2,3,5. 

We computed the Cohen’s Kappa agreement between the model outputs and the clinical scoring and found kappa values of (1) 0.80 and (2) 0.73.

\subsection{OncoNet can be applied to other anatomical regions in the scan}

To determine if OncoNet’s network architecture and training algorithm generalize to other anatomical regions of the exam, we evaluate OncoNet and the single pass ablation on a dataset of abdominal exams along with their associated SUVmax.
We find that OncoNet scores 0.8 [0.67-0.90] AUROC on treatment response prediction in the abdomen and pelvis with class-specific performance of 0.85 [0.72-0.97] AUROC on Progression, 0.74 [0.57-0.90] on Resolution and 0.80 [0.63-0.93] on Stable (See Fig \ref{fig:aperf}). Overall, OncoNet outperforms the single pass approach by 0.32 AUROC (p $<$ 0.01).

\begin{center}
\begin{table}
\begin{tabular}{ | c | c|c |c| c| } 

\hline
Model & AUROC & F1 & Precision & Recall \\ 
\hline
OncoNet & 0.80 [0.67-0.90] & 0.65 [0.49-0.80] & 0.67
[0.50-0.82] & 0.65
[0.51-0.78]\\ 
\hline
Single Pass & 0.48
[0.29-0.56] & 0.23 [0.11-0.36] & 0.20
[0.08-0.34]
 & 0.27
[0.14-0.43]
 \\ 
\hline

\end{tabular}
\caption{Class averaged performance metrics on the abdomen along with 95th percentile confidence intervals computed by bootstrap sampling over 1000 samples}
\end{table}
\end{center}

\begin{figure}
\includegraphics[width=\textwidth]{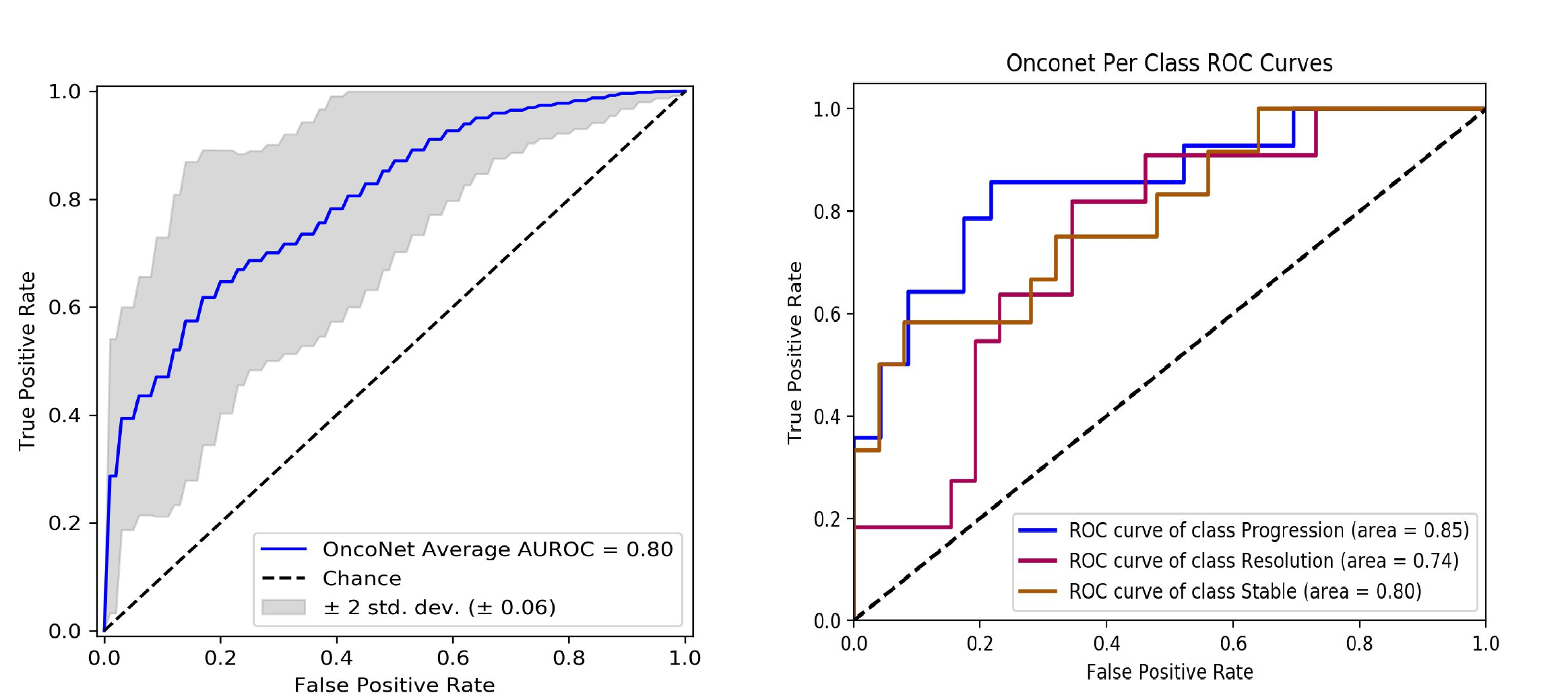}
\caption{ROC curves for OncoNet test performance on the abdomen. (Left) Microaveraged ROC curves over the three classes with the grey bounds indicating 2 standard deviations computed by bootstrap sampling on 1000 samples. (Right) The class-specific ROC curves for Progression (AUROC: 0.85), Resolution (AUROC:0.74) and Stable (AUROC: 0.80). }

\label{fig:aperf}
\end{figure}

}

{\spacing{1.4}
\section{Discussion}
The purpose of this study was to model the task of longitudinal treatment response prediction on multi-slice, multi-modality, multi-class oncologic imaging examinations to achieve automated determination of disease progression, improvement/response, or stability using pairs of FDG PET/CT studies obtained before and after treatment.  A siamese-style neural network (OncoNet) classified treatment response between consecutive PET/CT thoracic examinations with AUROC of 0.85 and generalized to an external dataset with AUROC 0.84. In addition, OncoNet achieved AUROC 0.80 when evaluating disease response in the abdomen. When mapping predictions to a common qualitative clinical scoring system for reporting change in disease over time (Deauville) the kappa agreement scores between OncoNet and subspecialist trained radiologist was 0.8, higher than reported literature between expert readers for the same task. Finally the labeled training dataset of 1954 imaging studies is made available to the open research and education communities to further innovate on this important problem. 

In the clinical interpretation of medical imaging, comparing consecutive medical image examinations is critical to evaluate disease severity and change over time and especially vital in oncologic imaging and management of therapy.  Prior work leveraging convolutional Siamese neural networks was shown to be successful in automated evaluation of disease severity and change over time in consecutive imaging studies in knee radiographs and retinal fundus images; this established the feasibility for comparing paired images from the same patient at two time points using a Siamese neural network allowing continuous measure of change between images without manual localization of the pathology of interest \cite{li2020siamese}. However this approach was limited as the medical imaging tasks in estimating only a binary output (same vs different) to achieve AUC 0.90 in evaluating knee osteoarthritis change and could accommodate only single 2-dimensional image examinations. To date there have been no published works exploring the automated quantitative comparative analysis of consecutive multi-slice multi-modality imaging examinations (i.e. CT, MR etc) such as the use case in this work with FDG PET/CT scans.  

Earlier work toward an end-to-end framework utilizing a weakly supervised approach to lesion detection and localization in PET/CT was found to be capable of excellent performance in leveraging an individual multi-slice imaging examination \cite{eyuboglu2021multi}. While this work represents innovation in automated analysis of FDG PET/CT using deep learning techniques, consideration of only the individual examination, without context to change over time in consecutive studies, ultimately lessens the clinical impact because comparative quantification of disease over time, especially during oncologic  therapy, is a chief indication for performing FDG PET/CT imaging.  Toward that goal,  there have been prior efforts toward automating the quantification of FDG-PET disease progression consisting of largely semi-automated approaches requiring significant manual input to achieve quantification \cite{lim2021quantitation}. For example the Auto-PERCIST software, based on traditional rules based software methodologies, can extract quantitative data for relevant imaging pathology (SUVmax, volume, etc) and was shown to reduce inter-reader variability between readers. However this system demands much of the interpreting physician, requiring user input for lesion identification, manual registration of comparison examinations, and human expert localization and selection of the reference tumor as the basis for comparison across studies. Other work used a CNN-based deep-learning approach to achieve automated segmentation of lung tumors in thoracic FDG-PET images based on phantom images or on small datasets without an end-to-end approach \cite{leung2018deep}.  By contrast our approach represents a fully automated end-to-end approach that reports progression, stable disease, or response without requiring any manual input from the human expert and achieving state of the art inter-rater agreement with human experts and is robust to external populations with varying scanner parameters, protocols, etc.

Prior work in automated disease progression have used class activation maps to visualize the progression of disease over multiple days for detecting COVID-19 from CT \cite{pu2021automated}. Other work on MRI to automatically assess treatment response primarily focused on using deep learning to segment tumor regions and using the intermediate extracted features to correlate to progression \cite{kickingereder2019automated}. Alternative approaches assigned severity scores to retinopathy imaging that were tracked longitudinally to determine response \cite{taylor2019monitoring}. Such approaches however need extensive segmentation annotations which are prohibitively expensive for modalities like PET/CT. They are also limited by not assessing the treatment response directly from data in an end to end fashion. OncoNet leverages pretraining for abnormality detection and derives supervision signal directly from associated radiology reports making it highly scalable and unbiased to hand crafted intermediate features.

FDG PET/CT has become indispensable in the routine clinical management of cancer patients and in therapeutic clinical trials \cite{silvestri2013methods, ettinger2017non, ravenel2014acr}. Response to cancer treatment is determined by serial size and SUV measurements of index cancerous lesions seen on PET/CT scans; the percentage of change in the measurements between scans is used to monitor response to therapy and demands standardized and reproducible assessments for meaningful comparisons and conclusions across multiple trials. For example the PET Response Criteria in Solid Tumors 1.0 (PERCIST 1.0) was  proposed in 2009 as a method to standardize the assessment of tumor response and includes the assessment of SUVmax on PET \cite{wahl2009recist, joo2016practical}.  But poor inter-reader agreement using scoring systems across examinations have been widely reported, as low as 0.14 (range 0.14-0.68) under a variety of experimental settings and comparison methods; the agreement rate is likely lower in clinical practice compared to ideal study settings \cite{kluge2016inter, weisman2020machine, kumar2013variance}. Such variability is an often cited hurdle to broader utilization of quantitative FDG PET/CT for response assessment especially in examining early treatment response-related changes \cite{ding2014pet, wahl1993metabolic}. While in practice the SUVmax is reasonably easy to determine with many forms of software, and as mentioned above can improve inter-reader variability across scans, but requires manual input.  OncoNet demonstrated automated agreement with a board certified radiologist with Cohen's kappa 0.8 and as the workflow requires no manual input, this approach could be used at scale for a variety of clinical and trial outcome determinations. 

Further, given proven value of FDG PET/CT in oncology and consequent continued increased in study volume, OncoNet could be used for rapid communication of high-level results to clinicians and patients, providing timely information regarding interval disease state to inform clinical decision making and easing clinical burdens; cancer patients are routinely asked to return to hospital-based imaging departments for FDG PET/CT imaging and for convenience schedule clinical appointments with an oncologist during the same visit. This leads to significant challenges as the oncologist requires knowledge of the disease state information from the imaging study, but study results may  not yet be available during the patient’s visit which leads to increased follow-up calls, calls to the nuclear imaging medical specialist , and patient anxiety waiting for results.  And while FDG PET/CT is intrinsically a quantitative imaging technique, in practice assessments of cancer response remain largely qualitative; thus many scoring systems have been developed as, for example, in lymphoma where quantitative PET data are converted into a five-point qualitative scale \cite{gallamini2014predictive}.  We found that OncoNet was at human-expert level agreement in both treatment response in a three point scale (i.e. progression, stable, response). Leveraging the routine use of OncoNet for simplified categorical measures of disease state results may lead to improved consistency and also help address the challenges of patient direct access to medical imaging results records as mandated under the final rule of the 21st Century Cures Act by providing simplified quantitative outcomes measures for tracking oncologic disease over time. 

This study includes several important limitations. This is a retrospective study design which comes with well-established shortcomings and inherent limitations. The deep learning model described was developed and trained on data from a single large academic institution and while robust external test evaluation was performed, additional study to comprehensively understand the generalizability of our model is needed to inform the direction of future work. The evaluation of this approach considered only a few tasks of the many use cases for FDG PET/CT, however, the methods and results should be considered when applying to other predictive tasks. Lastly, while our results are promising, delivering production-ready models in their final clinical form is beyond the scope of this study and additional work is needed before deploying such models in clinical practice. 

}

{\spacing{1.4}
\section{Conclusion}
In conclusion, this work describes the development of OncoNet as an end-to-end approach for quantitatively determining longitudinal treatment response assessment on multi-slice multi-modality oncologic FDG PET/CT imaging examinations. OncoNet achieved an AUROC of 0.85 on automated determination of disease resolution, stability or progression  using pairs of FDG PET/CT studies obtained before and after treatment with robust external validation (AUROC 0.84). OncoNet further achieves agreement with a board certified radiologist with a kappa of 0.8. OncoNet's methodology and associated annotated dataset are designed to achieve automated quantitative oncologic imaging evaluation over time with potential broad implications for cancer care and contributes to the broader machine learning in healthcare research community. 

}

\clearpage
\restoregeometry

\paragraph{Acknowledgements.} 

We would like to acknowledge the GE Blue Sky team (Elizabeth Philps, Omri Ziv, Gil Kovalsky, Melissa Desnoyers, Shai Kremer) for their financial support for this industry-academic collaboration.

\paragraph{Author information.} 

Matthew P Lungren is a visiting researcher at Microsoft and provides consulting services to Philips, Segmed, Centaur, Bunker Hill, and Nines Radiology; received research funding for this work from GE Healthcare and a research grant from the National Library of Medicine of the NIH.

A. S. Chaudhari has provided consulting services to SkopeMR, Inc., Subtle Medical, Chondrometrics GmbH, Image Analysis Group, Edge Analytics, ICM, and Culvert Engineering; is a shareholder of Subtle Medical, LVIS Corporation, and Brain Key; and receives research support from GE Healthcare and Philips.

The authors declare no conflict of interest. Correspondence should be addressed to anirudhjoshi@cs.stanford.edu.


\clearpage

\section*{References}
{
\spacing{0.85}
\bibliographystyle{naturemag}
\bibliography{refs}
}

\end{document}